\documentclass[cite]{PoS}
\usepackage{amssymb}

\newcommand{\dd}[1]{\mbox{d} #1}

\newcommand{\dddd}[1]{\dd{{}^{4} #1}}

\newcommand{\tr}{\mbox{tr}}

\newcommand{\sgnc}{{\rm sgn}_{\mathcal{C}}}

\newcommand{\deltac}{\delta_{\mathcal{C}}}
\newcommand{\intc}{\int\limits_{\mathcal{C}}}
\newcommand{\refeqs}[2]{Eqs.~(\ref{#1}-\ref{#2})}
\newcommand{\reffig}[1]{Fig.~\ref{#1}}
\newcommand{\refeq}[1]{Eq.~(\ref{#1})}

\newcommand{\prn}[1]{\left(#1\right)}
\newcommand{\cG}{\mathcal{G}}
\newcommand{\Gammaint}{\Gamma_{int}}

\newcommand{\twoPI}{_{\rm 2PI}}

\newcommand\ihalf{\frac i2}
\newcommand{\bp}{\vec p}

\title{Nonequilibrium field theory from the 2PI effective action}

\ShortTitle{Nonequilibrium field theory}
\author{\speaker{Szabolcs Bors\'{a}nyi}\\
Institute of Theoretical Physics, \\
Philosophenweg 16, Heidelberg 69120, Germany\\
E-mail: \email{s.borsanyi@thphys.uni-heidelberg.de}}

\abstract{
Nonperturbative approximation schemes are inevitable even in weakly
coupled theories if the nonequilibrium behavior of quantum fields
is investigated. The two-particle irreducible (2PI) effective action
formalism provides an efficient framework for obtaining resummation schemes
both in and out of equilibrium.  We briefly review the these techniques
and discuss recent findings for nonequilibrium field theories. 
}

\FullConference{29th Johns Hopkins Workshop on current problems
in particle theory: Strong matter in the heavens,\\
Budapest, Hungary\\
August, 1-3 2005}

\begin{document}

\section{Introduction}

The new observations of the heavy ion collider facilities represent a great
challenge to particle field theory. The achieved volume and life time of
the produced hot and dense matter may now enable thermalization.
On the other hand, due to the developments on the theory side,
calculations of nonequilibrium processes become increasingly feasible.

It has been recently observed that the phenomenological description of
particle flow based on ideal hydrodynamics is extremely successful
at RHIC energies \cite{v2hydro}. The used models assume local equilibrium
already before 1 fm/c. The competence of ideal hydrodynamics indicates a very
early thermalization of the produced plasma \cite{Heinz:2004pj}.  A further
signal of an at least
partial equilibration is that the global abundances from collision experiments
obey simple statistical models assuming chemical equilibrium \cite{chemeq}.

The observed early onset of equilibrium calls for a novel theoretical
paradigm. The perturbative thermalization time estimates are far beyond
the experimental expectations. As a first step one has to understand, what
level of equilibration is required to interpret the experimental data.
The success of ideal hydrodynamical description might be explained by the early
onset of an isotropic equation of state \cite{isovsthermal} that may be present
even in a far-from-equilibrium quantum field \cite{prethermalization}.  Then
one seeks for explosive processes that drive the system towards isotropy and
equilibrium.  A promising candidate is the development of plasma instabilities
that may dramatically accelerate the evolution of anisotropic fields
\cite{plasmainstabilities1,plasmainstabilities2}.

Nonequilibrium methods are also asked for in early Universe scenarios.
Much of the efforts have been put into the quantitative understanding of
the theory of reheating, which explains the rapid re-population
of the dilute Universe after inflation \cite{reheating}. The final temperature
is an important input to models of baryogenesis \cite{noneqbaryogenesis}.

The intriguing questions of non-thermal field evolution cannot be answered by a
simple perturbative analysis. Even in a weakly coupled theory high orders
become relevant with the evolution of the time. The elapsed time appears next
to the coupling constant, making the effective coupling arbitrarily big at late
times.  This phenomenon, called \textit{secularity}, invalidates approximations
that truncate higher order diagrams in the equation of motion of the Green's
function \cite{secularity}. A way to evade this trap is to use self-consistent
schemes \cite{Voskresensky}.

There has been important progress in our understanding of nonequilibrium
quantum fields using suitable resummation techniques based on
two-particles-irreducible (2PI) generating functionals~\cite{Berges:2004yj}.
It has been shown by Cox and Berges \cite{CoxBerges} that the
Kadanoff-Baym equations are suitable for direct numerical treatment. They used
the 2PI effective action formalism \cite{CJT} to calculate the self energies
in as systematic way.  They found that the studied scalar field reaches thermal
equilibrium even in 1+1 dimensions, where binary collision are
forbidden by kinematics on the level of the Boltzmann equation.

The 2PI effective action techniques (or $\Phi$ derivable
approximations) are known to provide time reversal symmetric,
energy conserving equations for the field propagator \cite{Voskresensky}.
The equations involve the resummation of those diagrams that carry the time
in the effective coupling, this cures the secularity of the perturbative
treatment.

The 2PI equations of motion have been solved for various models since the
pioneering work by Cox and Berges \cite{CoxBerges}: ranging from the simplest
$\Phi^4$ scalar theory \cite{AartsBergesSpectral, Arrizabalaga2,Isotropization}
through the scalar O($N$) theory with next--to--leading order large $N$
resummation \cite{BergesON,BielefeldHeidelberg,
AartsBergesClassical,BergesSerreau,Cooper,Arrizabalaga1} to the chiral meson
model \cite{Fermions}, often considered as a prototype of QCD.

\begin{figure}
\begin{center}
\hbox{
\includegraphics[width=7cm]{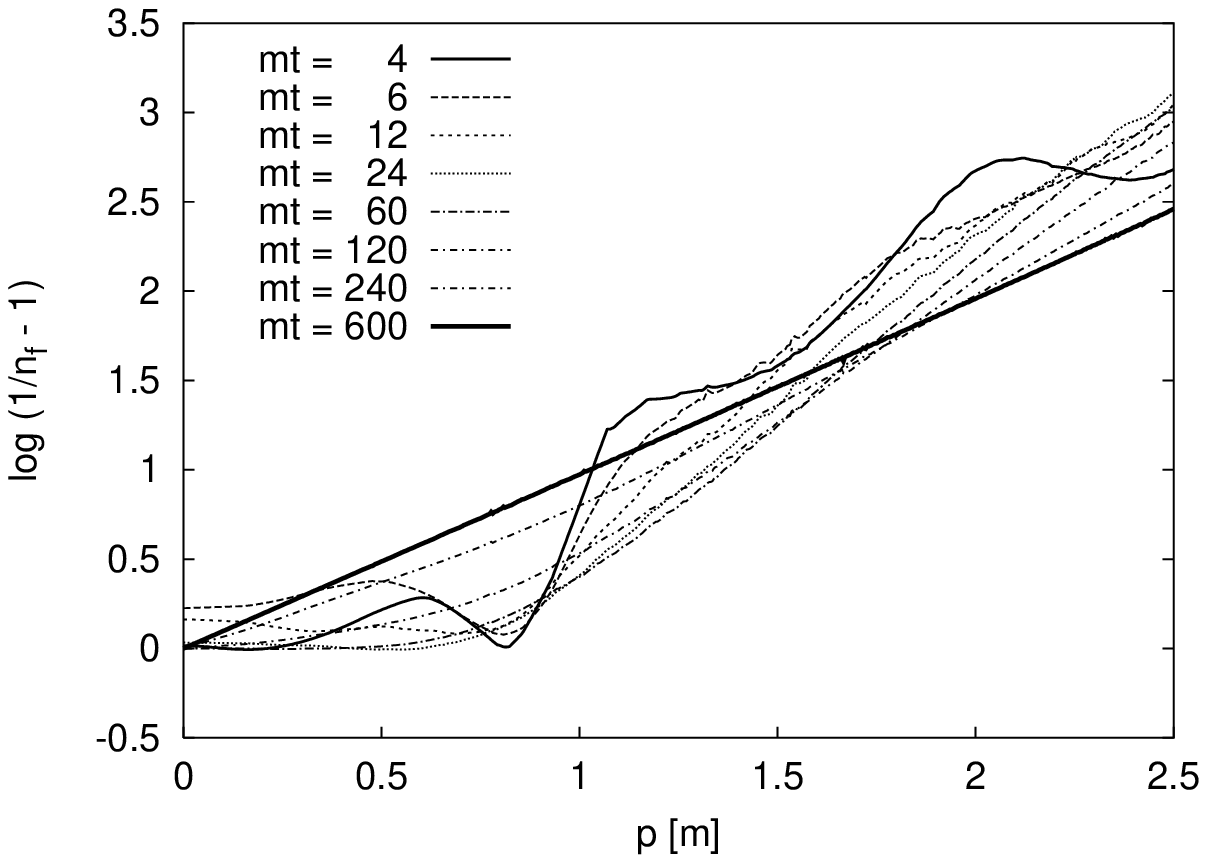}
\hspace{5mm}
\includegraphics[width=7cm]{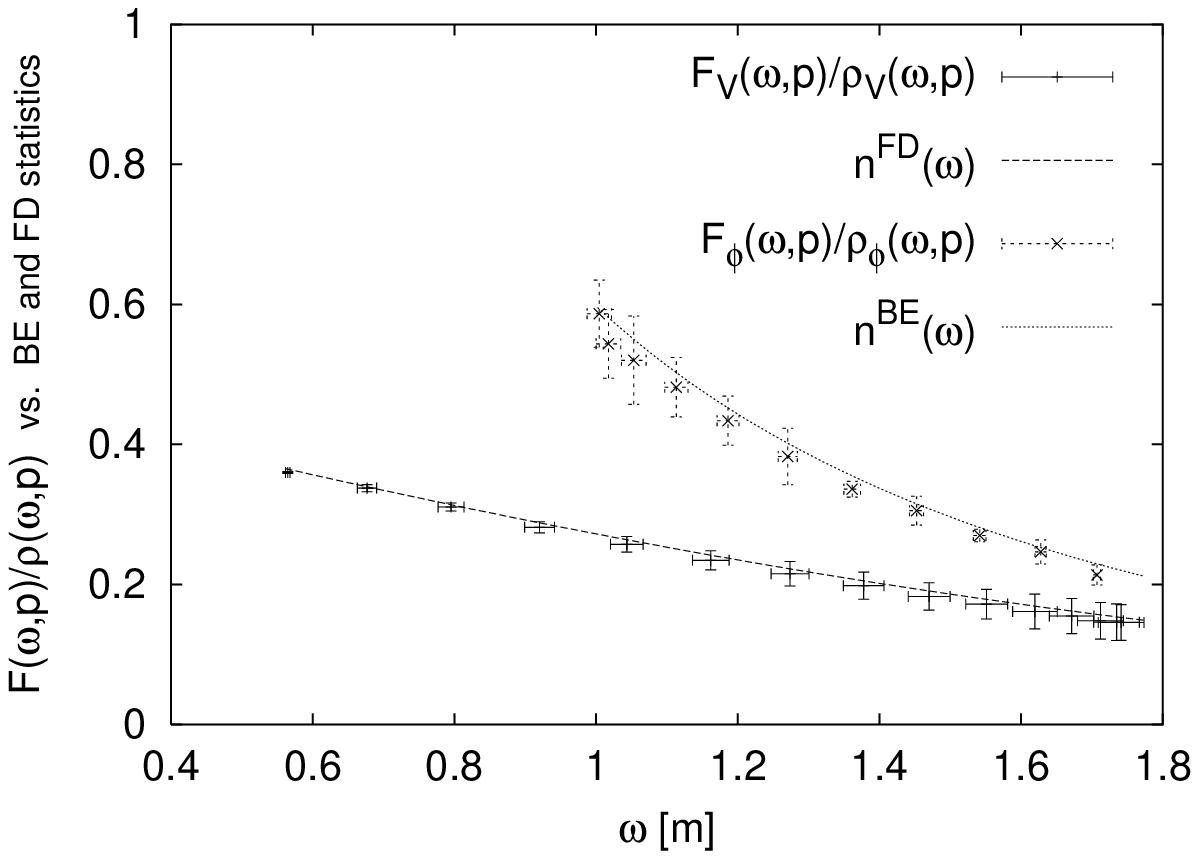}
}
\end{center}
\caption{\label{fig:BEFD}
Evolution of the particle distribution function in the chiral quark model
\cite{Fermions}.  To the left, the fermionic particle spectrum is shown for
various times.  The spectra have been transformed so that a linear
corresponds to the Fermi-Dirac statistics. To the right, the final
distribution is displayed without the transformation. The coupled system of
fermions and scalars dynamically realizes the Fermi-Dirac and Bose-Einstein statistics.
}
\end{figure}

The 2PI formalism was used to show for the first time from first
principles the formation of the Bose-Einstein and Fermi-Dirac statistics
in a system of coupled fermions and scalars \cite{Fermions}. The used equations
do not involve any of the statistical factors or constraints for the occupation
numbers, not even the concept of particles. The resulting distributions are
purely an outcome of the coupled field dynamics (See Fig.~\ref{fig:BEFD}).

In the next sections we first review the 2PI effective action formalism 
on the example of a scalar theory. Then we discuss the various resummations
necessary to calculate the self-consistent $n$-point functions to the desired
accuracy.  The 2PI results have been checked both in equilibrium and out of
equilibrium, we briefly mention the main results of these tests.
We also highlight some of the
numerical results on the basic time scales of nonequilibrium evolution.

\section{The formalism}

The objective is to give an equation for the propagator and also for the higher
$n$-point functions of an evolving field.  The 2PI formalism does not use the
concept of quasi-particles, it establishes the relationship
between various $n$-point functions in a systematic truncation scheme.
To keep this review simple,
we restrict our discussion to the scalar propagator, which already gives
account for the evolution of particle distribution. 
The equations are worked out for higher $n$-point functions in
Ref.~\cite{Julien}. The 2PI formalism for multiple scalar fields
is discussed in detail in Refs.~\cite{BielefeldHeidelberg,LA2PI}, for
fermionic fields it has been elaborated in Refs.~\cite{Fermions,UrkoYukawa}.
For renormalizable theories the 2PI equations of motion are also
renormalizable, the correct way to define a finite theory has been discussed
by many authors \cite{vanHeesRenorm,UrkoRenorm,Julien,LA2PI,UrkoYukawa}.
The interested reader is referred to the in-depth review in
Ref.~\cite{Berges:2004yj}. 

The proper two-point function, that we are looking for, is defined in terms
of the one-particle irreducible (1PI) effective action:
\begin{equation}
\cG^{-1}(x,y)=-i\frac{\delta^2 \Gamma[\phi] }{
\delta\phi(x)\delta\phi(y)}
\label{properG}
\end{equation}
In the case of more than one field components $\cG$ is a matrix in the
field indices as well as in the space-time coordinates.

The $\Gamma[\phi]$ effective action is now a functional of the space and time
dependent background field. A straightforward approximation would be a
coupling expansion in $\Gamma[\phi]$. A danger in this
approach is, that a Taylor expansion in $\phi$ simply cancels the
higher proper $n$-point functions. Negative results warn us that
it is required to include high powers of $\phi$ in the effective action.
This can be achieved by imposing the truncation to the Legendre transform of
$\Gamma[\phi]$. This way we selectively resum 1PI diagrams to infinite order.

The two-particle irreducible (2PI) effective action is defined as a Legendre
transform of the effective action with respect to the source $K(x,y)$ of the
two-point composite field operator ($\phi(x)\phi(y)$):
\begin{equation}
\Gamma_{\rm 2PI}\left[\phi, \cG[\phi,K]\right]
= \Gamma[\phi] - \frac12 \intc\dddd{x}\intc\dddd{y}
\left(
\cG[\phi,K](x,y)K(x,y)+\phi(x)K(x,y)\phi(y)
\right)\, .
\end{equation}
The subscript $\mathcal{C}$ refers to the closed time path (CTP) contour for
the time integral. 

The second argument of the 2PI effective action $\Gamma\twoPI[\phi,G]$
is a generic two-point function $G(x,y)$. For any given $\phi$ background 
and vanishing sources %the propagator $\cG$ extremizes $\Gamma\twoPI$:
%\begin{equation} \label{eq5}
%  \frac{\delta \Gamma_{\rm 2PI}\left[ \phi, G \right]}{\delta \phi_a \left( x \right)} = 0 \;,
%\end{equation}
\begin{equation} \label{eq6}
  \frac{\delta \Gamma_{\rm 2PI}\left[ \phi, G \right]}{\delta G \left( x, y \right)} = 0 \;,
\end{equation}
with $G=\cG[\phi]$. From $\Gamma\twoPI$ it is straightforward to go back to the
1PI effective action using the identity 
$\Gamma[\phi]=\Gamma\twoPI[\phi,\cG[\phi]]$.

Unfortunately, the exact $\Gamma_{\rm 2PI}$ functional is not known.
Following Cornwall, Jackiw and Tombulis \cite{CJT}
one can decompose the 2PI effective action into tree-level (1st term),
one-loop level (2nd and 3rd term) and higher order (4th term) contributions.
\begin{eqnarray}
 \Gamma_{\rm 2PI}\left[ \phi, G \right] 
  & = & S \left[ \phi \right] + \frac{i}{2} \tr_{\mathcal{C}} \left[ \log \left[ G^{-1} \right] \right] + \frac{i}{2} \tr_{\mathcal{C}} \left[ G_0^{-1}[\phi] G \right] \label{eqdorig} \\
  &   & + \Gamma_2 \left[ \phi, G \right] + const \nonumber \;.
\end{eqnarray}
Here $G_0[\phi]$ denotes the free propagator on the background, and $S$ is
the classical action. The decomposition (\ref{eqdorig})
defines a term $\Gamma_2[\phi,G]$ that can be entirely represented by
two-particle irreducible diagrams, i.e. these diagrams do not fall apart
if any two of its lines are cut.

We will use a slightly modified decomposition, where the background dependence
of the free propagator is put into the interaction part
$\Gamma_{\rm int}[\phi,G]$:
\begin{eqnarray}
 \Gamma_{\rm 2PI}\left[ \phi, G \right] 
  & = & S \left[ \phi \right] + \frac{i}{2} \tr_{\mathcal{C}} \left[ \log \left[ G^{-1} \right] \right] + \frac{i}{2} \tr_{\mathcal{C}} 
\left[ G_0^{-1}[\phi\equiv0] G \right] \label{eq8} \\
  &   & + \Gamma_{\rm int} \left[ \phi, G \right] + const \nonumber \;.
\end{eqnarray}

In the 2PI formalism the truncation is realized by the formal expansion of 
$\Gamma_{\rm int}[\phi,G]$ in the coupling constant or in an
other small parameter, such as the inverse number of field components.
There are Feynman rules to construct $\Gammaint$ from the selected diagrams.
The most important ingredient of these rules is to write a $G(x,y)$ function
for each propagator. For the vertices, however, the bare couplings
appear (with counterterms, eventually).

For the most trivial case of scalar $\lambda\Phi^4/24$ theory one simply keeps:
\begin{eqnarray}
\Gamma_{\rm int}[G]=
-\frac{\lambda}4\intc dx\, \phi(x) G(x,x) \phi(x)
-\frac{\lambda}8\intc dx\, G^2(x,x) \nonumber\\
+ i\frac{\lambda^2}{12} \intc\intc dx dy\, \phi(x)G^3(x,y)\phi(y)
+ i\frac{\lambda^2}{48} \intc\intc dx dy\, G^4(x,y)\, .
\label{2piphi4}
\end{eqnarray}
The first two terms correspond to the (collisionless) Hartree approximation
\cite{hartree}, the next two terms give account for the ``particle
collisions'' at lowest order. An expansion to higher orders in $\lambda$
is found in Ref.~\cite{Arrizabalaga2}. 

When using a truncated $\Gamma\twoPI$ we have to rethink the exact identities
derived so far. We define the resummed two-point function $\bar
G[\phi](x,y)$ as the solution of \refeq{eq6} with the truncated $\Gamma\twoPI$.
Using this as an argument of $\Gamma\twoPI$ one uses
\begin{equation}
\Gamma[\phi]=\Gamma\twoPI\left[\phi,\bar G[\phi]\right]
\end{equation}
as a definition for the truncated 1PI effective action \cite{vanHeesSymmetry}.
As already noted,
this relation turns to an identity in the case without truncation.
This truncated $\Gamma[\phi]$ is used then to define $\cG$ according to
\refeq{properG}. This $\cG(x,y)$ propagator is in general, not identical to
$\bar G[\phi](x,y)$

As a first step we have to solve \refeq{eq6}, which can be put to the form
of a Schwinger-Dyson equation using \refeq{eq8}:
\begin{eqnarray} \label{eq73}
 \overline{G}^{-1} \left( x, y \right) = 
G_{0}^{-1}[\phi\equiv0] \left( x,y \right)
  - \Sigma\left[\overline{G}[\phi]\right] \left( x, y \right)
\end{eqnarray}
with
\begin{eqnarray} \label{eq44}
  \Sigma[\overline{G}] \left( x, y \right) &=& 2 i
\frac{\delta \Gamma_{\rm int} \left[ \overline{G} \right]}
{\delta \overline{G} \left( x, y \right)}\,.
\end{eqnarray}
Here the self energy ($\Sigma$) is built of one-particle irreducible diagrams
constructed form resummed lines ($\bar G[\phi]$) and bare vertices.
(Coming from the first term in \refeq{2piphi4} it may actually contain
a diagram $\sim\deltac(x,y)\phi(x)\phi(y)$, which is not 1PI. Using the
original decomposition (\ref{eqdorig}) this term would appear
in $G_0^{-1}[\phi]$.)

In the second step we evaluate the 2PI effective action at $\bar G[\phi]$ 
and calculate $\cG$ as follows:
\begin{equation}
{\cal G}^{-1}[\phi](x,y)=-i\frac{\delta^2 \Gamma[\phi] }{
\delta\phi(x)\delta\phi(y)}
=
-i\frac{\delta^2 \Gamma_{\rm 2PI}[\phi,\overline{G}[\phi]] }{
\delta\phi(x)\delta\phi(y)}\,.
\label{properG2}
\end{equation}
Analogously, any higher $n$-point functions are available if
$\overline{G}[\phi]$ is known for any $\phi$ background. 

In the symmetric phase ($\phi\equiv0$) \refeq{properG2} takes
a simpler form if the model has the $Z_2$ ($\phi\leftrightarrow -\phi$)
symmetry:
\begin{equation}
{\cal G}^{-1}[\phi\equiv0](x,y)=
G_0^{-1}[\phi\equiv0](x,y)
-i\left.
\frac{\delta^2 \Gamma_{\rm int}[\phi,\overline{G}[\phi\equiv0]] }{
\delta\phi(x)\delta\phi(y)}
\right|_{\phi\equiv0} \, .  \label{properGsymm}
\end{equation}
We emphasize at this point, that in the broken phase additional terms appear
\cite{Julien}. Without these additional terms the propagator $\cG$ would
violate Goldstone theorem \cite{vanHeesSymmetry,BielefeldHeidelberg}. 

There are some ``friendly'' truncations (including \refeq{2piphi4})
for which $\Gamma_{\rm int}$ has the symmetry:
\begin{equation}
\left.\frac{\delta^2 \Gamma_{\rm int}[\phi,G]}{
\delta\phi(x)\delta\phi(y)}\right|_{\phi\equiv0}
=
\left.2\frac{\delta^2 \Gamma_{\rm int}[\phi,G] }{
\delta G(x,y)}\right|_{\phi\equiv0}\,.
\label{friendly}
\end{equation}
This further simplifies \refeq{properGsymm} to
\begin{equation}
{\cal G}[\phi\equiv0](x,y)=\overline{G}[\phi\equiv0](x,y)\,.
\end{equation}
For a $Z_2$ symmetric exact theory \refeq{friendly} is always granted,
as expected, $\cG$ and $\overline{G}$ are then equivalent. 

All the models numerically investigated so far
(the three-loop $\Phi^4$ theory, the O($N$) model truncated
at next-to-leading order in $1/N$ and the two-loop chiral quark model)
have this nice feature. This explains why numerical works directly
use the solution of the Schwinger-Dyson equation (\ref{eq73}).

\section{Evolution equations}

\refeq{eq73} uses complex and nonanalytic functions on CTP contour,
thus, it is not suitable for a numerical treatment.
Therefore we introduce the
statistical propagator $F(x,y)$ and the spectral function $\rho(x,y)$
and the corresponding functions for the self energy,
too \cite{AartsBergesSpectral}:
\begin{eqnarray}
\overline{G}(x,y)&=&F(x,y)-\frac{i}2\rho(x,y)\sgnc(x_0,y_0)\,,
\label{Gdecomposition}\\
\Sigma[\overline{G}](x,y)&=&-i\Sigma_{0}(x)\deltac(x,y)
+\Sigma^F(x,y)-\ihalf\Sigma^{\rho}(x,y)\sgnc(x_0,y_0)\,.
\end{eqnarray}
The real ($F$) and imaginary ($\rho$) part capture different physical
information: $\rho$ gives what states are available and how stable they are,
while $F$ tells how they are occupied. In equilibrium, the $F$ and $\rho$
functions are connected by the KMS condition, but they are a priori
independent out of equilibrium.
These functions do not depend on contour variables, all discontinuities are
encapsulated in the contour delta $\deltac$ and sign function $\sgnc$.
One can transform \refeq{eq73} into integral equations, where
the integral is defined on the closed-time-path contour \cite{CTP}.
Using the $F$ and $\rho$ functions in these integral equations
the discontinuities appear as boundaries of the time integral.
For a scalar field theory, the equations of motion in
terms of the $F$ and $\rho$ functions read:
\begin{eqnarray}
\prn{\partial_x^2+m^2+\Sigma_{0}(x)}F(x,y)&=&
\int\limits^{y_0}_{t_0} dz\Sigma^F(x,z)\rho(z,y)
-\int\limits^{x_0}_{t_0} dz\Sigma^\rho(x,z)F(z,y)\,,
\label{Feom}
\\
\prn{\partial_x^2+m^2+\Sigma_{0}(x)}\rho(x,y)&=&
\int\limits^{y_0}_{x_0} dz\Sigma^\rho(x,z)\rho(z,y)\,.
\label{rhoeom}
\end{eqnarray}
These equations are exact, the approximation is encoded into the
used self energy, given by \refeq{eq44}. Working out the self energy
for the model of choice we can close \refeqs{Feom}{rhoeom}.
For the one-component scalar $\lambda\Phi^4/24$ theory at
three-loop level two diagrams contribute (at $\phi\equiv0$)
\cite{AartsBergesSpectral} 
\begin{eqnarray}
\Sigma^F(x,y)&=&-\frac{\lambda^2}6F(x,y)\left(F^2(x,y)-\frac{3}{4}\rho^2(x,y)\right)\,,\\
\Sigma^F(x,y)&=&-\frac{\lambda^2}2\rho(x,y)\left(F^2(x,y)-\frac{1}{12}\rho^2(x,y)\right)\,.\\
\Sigma_0(x)&=&\frac{\lambda}{2}F(x,x)\,.
\end{eqnarray}

The solution of the resulting closed set of equations is not accessible without
numerical methods. If the initial ensemble is spatially homogeneous, the
two-point functions depend on five variables ($x_0,y_0,\vec x-\vec y$) only.

The actual form of the equations of motion (\ref{Feom}-\ref{rhoeom}) 
depend on the type of the initial conditions. 
The most simple choice is to assume a known
Gaussian density operator at initial time $t_0$, the presented form of
Eqs.~(\ref{Feom}-\ref{rhoeom}) reflect this choice.
In this case the integrals start form $t_0$, and the numerical
integration of \refeqs{Feom}{rhoeom} is conceptually straightforward
(but may be technically involved).

The initial conditions for the second order differential equations 
can be given by setting $F(x_0=t_0,\vec x;y_0=t_0,\vec y)$ and its
first order time-derivatives. The $\rho(x_0=t_0,\vec x;y_0=t_0,\vec y)$
function is zero identically, its derivative is constrained by Heisenberg`s
commutation relation:
$\partial_{x_0}\rho(x_0=t_0,\vec x;y_0=t_0,\vec y) =\delta(\vec x-\vec y)$.

An initial particle distribution $n_0(\bp)$
 can be easily prescribed using Gaussian initial conditions:
\begin{eqnarray}
F(x^0,y^0;\bp)|_{x^0=y^0=t_0} 
&=& \frac{n_0(\bp)+1/2}{\omega_p} \, ,
\label{eq:init1}\\
\partial_{x^0}\partial_{y^0}F(x^0,y^0;\bp)|_{x^0=y^0=t_0} &=& 
[n_0(\bp)+1/2]\omega_p\,,
\label{eq:init2}\\
\partial_{x^0}F(x^0,0;\bp)|_{x^0=t_0} = 0\,,  
\end{eqnarray}
with $\omega_p \equiv \sqrt{{\bp}^2 + M_0^2}$ and $M^2_0$
is the initial time physical mass. Here the propagators are
Fourier transformed in space with respect to the relative coordinate
$\vec x-\vec y$. 

\section{\label{sec:resummation}The resummation scheme}

\begin{figure}
\begin{center}
\includegraphics[width=15cm]{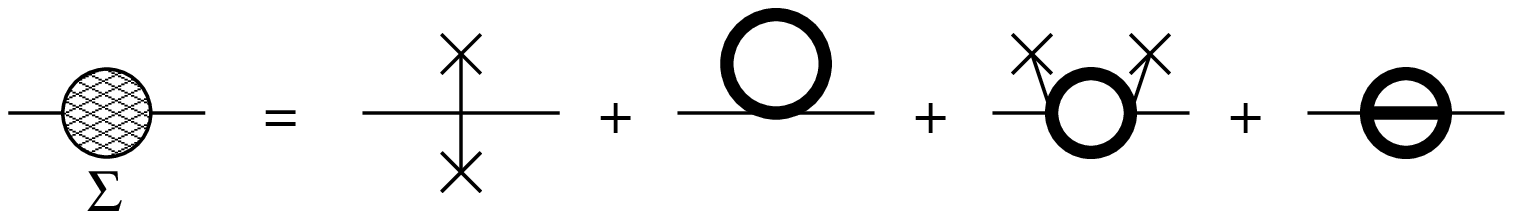}\par
\includegraphics[width=15cm]{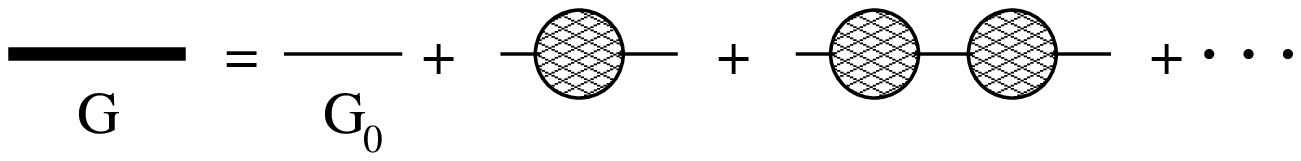}
\end{center}
\caption{\label{fig:sigmaresum}
Graphical representation of the 2PI equations of motion. The upper equation
shows the self energy in the simple truncation of the scalar $\Phi^4$ theory
at three loop order.
The thick lines stand for the self-consistent propagator ($G$),
which is a subject
of Dyson resummation (lower equation) using the free propagator (thin lines) and
this self energy.
} \end{figure}

Let us now review the resummations automatized by the 2PI formalism
on the simple example of a scalar $\Phi^4$ theory. The considered
approximation we define by giving the 2PI effective action explicitely
in \refeq{2piphi4}. The corresponding
self energy equation (\ref{eq44}) can be graphically expressed as
seen in \reffig{fig:sigmaresum}. The first two diagrams contribute
at leading order to the self energy\footnote{
In the broken phase the external field ($\Phi$) may carry inverse powers of the
coupling, this will promote the $\Phi$-dependent diagrams to lower orders.}.
The next two terms account for the next-to-leading order effects.

\begin{figure}
\begin{center}
\includegraphics[width=8cm]{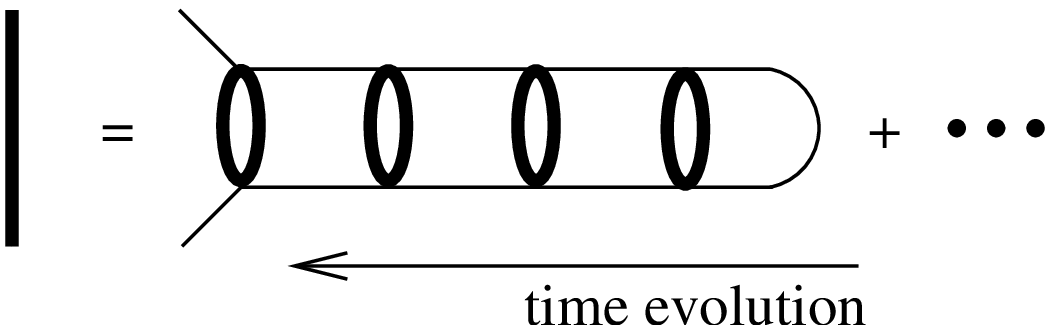}
\end{center}
\caption{\label{fig:scalarladder}
Nonequilibrium interpretation of the ladder resummation. For any perturbative
order there is a time when its contribution starts being relevant. A given
order may account for some finite number of elementary collisions. After
that number of elementary processes new orders will become of the same order.
The 2PI ladder resummation considers all orders and solves the problem
of secularity.
}
\end{figure}

By unrolling these equations and drawing the diagrams contributing to $G$
in terms of the free (or initial time) propagator one reveals the
elementary physical processes caught by this scheme. As shown in
\reffig{fig:scalarladder} a typical diagram is a ladder with finite number
of rungs. Thinking of quasi-particles,
each of these rungs represents an elementary scattering in the
time dependent bath of other particles. One may associate a time scale $\tau$
to these elementary processes. In a period of $n\tau$,
$n$ elementary collisions occur per 
particle on average. Beyond this period of time
diagrams with more rungs start becoming relevant, hence 
higher perturbative orders are necessary. This phenomenon, called secularity,
forbids the perturbative treatment of nonequilibrium fields, and calls
for the resummation of this chain of ladders. This chain of ladders is
what 2PI actually resums.

\begin{figure}
\begin{center}
\includegraphics[width=10cm]{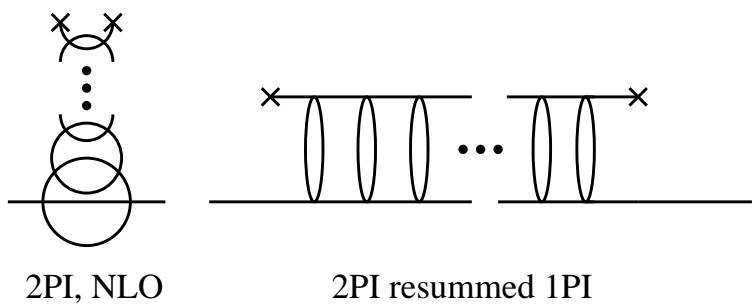}
\end{center}
\caption{\label{fig:2pi1pi}
Graphs contributing to the $\phi$-dependent part of the 2PI-resummed propagator
at next-to-leading order. The 2PI equation of motion resums a ladder in one
channel (left hand side).  To the proper two-point function ${\cal G}$,
however, additional diagrams contribute (right hand side), that play an important
role in the broken phase: they make sure that the Goldstone theorem is
fulfilled. The missing diagrams can be calculated by solving the Bethe-Salpeter
equation.
}
\end{figure}

\begin{figure}
\begin{center}
\includegraphics[width=8cm]{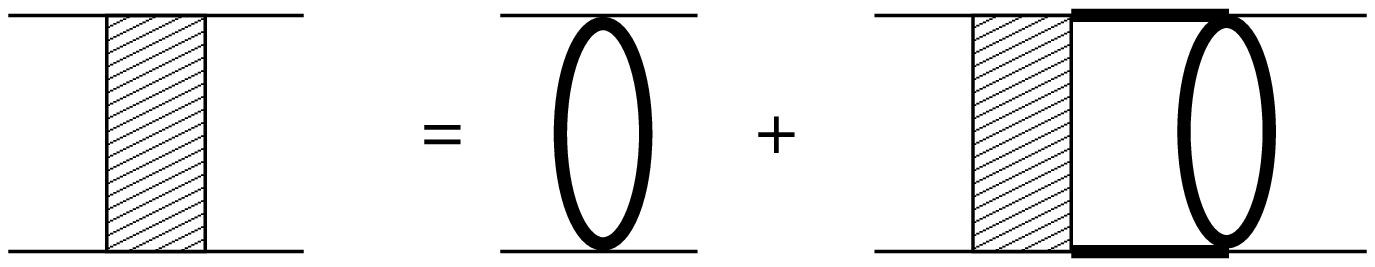}
\end{center}
\caption{\label{fig:bs}
Graphical representation of the Bethe--Salpeter equation. The thick
lines stand for the solution of the propagator equation at vanishing
background. 
}
\end{figure}

If one now tries to calculate a four-point function out of $\Sigma[\bar G]$ 
by cutting a
line, one realizes soon that only one of the three possible channels
are resummed, the Bose symmetry is broken. This is the reason we
did not define the propagator as $\bar G$, the solution of the 2PI equations of
motion, but used the 1PI effective action instead to introduce
the proper two-point function ${\cal G}$ in \refeq{properG}.
The difference between $\bar G$ and ${\cal G}$ are diagrammatically exemplified
in \reffig{fig:2pi1pi}. In many cases (see \refeq{friendly}) they agree
at $\phi\equiv0$, therefore we show a selection of $\phi$-dependent diagrams only.
The channels missing from $\bar G$ can be included by solving
a further (Bethe--Salpeter) equation \cite{vanHeesRenorm,UrkoRenorm,Julien}.
This equation follows from the definition of ${\cal G}$ without additional
theoretical input.
In \reffig{fig:bs} we give its diagrammatic form, showing the NLO contribution
only.  One can show for the truncations obeying \refeq{friendly}, that the
Bethe--Salpeter equation is the equation of motion for ${\delta^2 \bar
G(\phi;x,y)}/{\delta\phi^2}$ at vanishing background, which is required to
carry out the derivatives in \refeq{properG}. The four-point box contributes
symmetrically in each channel to \refeq{properG}, this restores the Bose
symmetry \cite{vanHeesSymmetry,Julien}.

\begin{figure}
\begin{center}
\hbox{
\vbox{\hsize=6cm
\includegraphics[width=6cm]{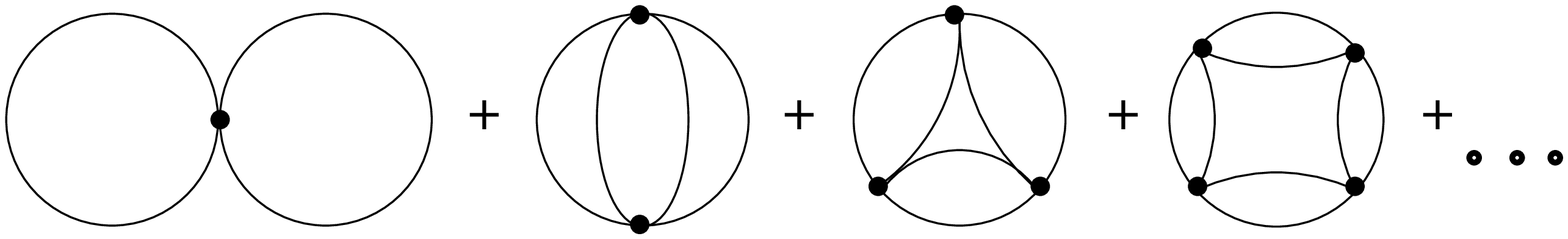}\par
\includegraphics[width=5cm]{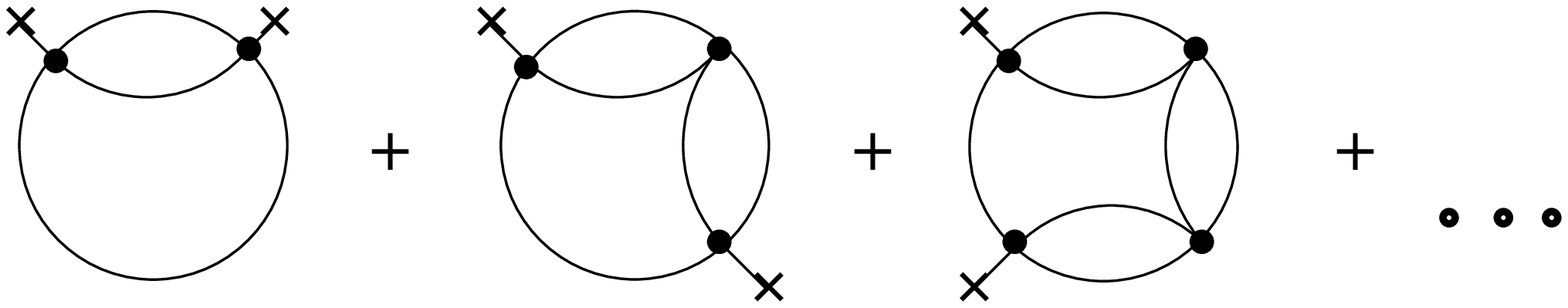}
}
\hskip 5mm
\vrule
\hskip 5mm
\vbox{\hsize=6cm
\includegraphics[width=6cm]{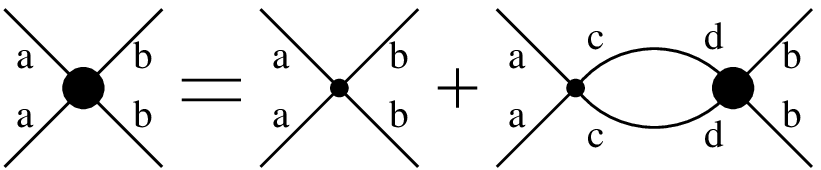}\par
\includegraphics[width=6cm]{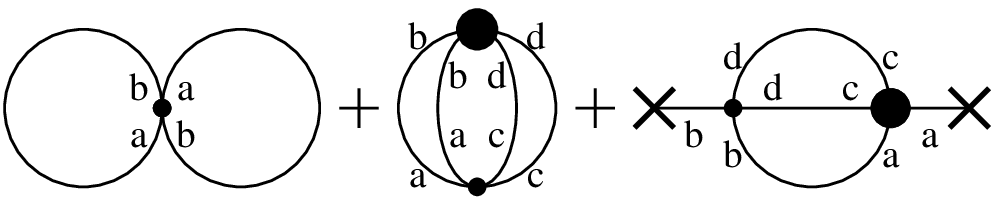}
}
}
\end{center}
\caption{\label{fig:onresum}
Large-$N$ resummation in the O($N$) scalar theory. The 2PI effective action
to next-to-leading order in $1/N$ consists of an infinite series of diagrams
(left panel). These can be resummed by introducing a self-consistent coupling
(big dot on the right panel) \cite{BielefeldHeidelberg}.
}
\end{figure}

The topology of the resummed diagrams become more complicated for the O($N$)
symmetric scalar theory in the large $N$ expansion.  As detailed in
Refs.~\cite{BergesON,BielefeldHeidelberg}, an additional resummation is
required on the level of the 2PI effective action.  Then $\Gamma_{\rm
int}[\phi,G]$ is already an infinite series of diagrams \reffig{fig:onresum}
(left). By introducing a self-consistent vertex function (see
\reffig{fig:onresum} (right-top)) one may reduce the 2PI effective action to a
sum of finite number of 2PI diagrams \reffig{fig:onresum} (right-bottom). The
concept of self-consistent vertex functions can be generalized and made
systematic in the context of the $n$-particle irreducible effective actions
\cite{npi}.

This large-$N$ resummation is always
necessary if the occupation numbers are large (proportional to the inverse coupling). Any additional loop in \reffig{fig:onresum} (left) carries additional
powers of occupation numbers that compensate the coupling. This typically
happens in preheating dynamics of the early universe \cite{BergesSerreau}.

2PI can be also used to derive Boltzmann equations. From the 2PI
effective action in \refeq{2piphi4} one arrives to a kinetic theory that
accounts for the $2\to 2$ scattering of quasi-particles \cite{CalzettaHu}.
The connection of the 2PI equations with kinetic theories justifies the
interpretation of the ladder resummation above, and also makes clear what
additional assumptions are made in such theories. Since 2PI does not make
these assumptions, it can be used to establish the range of validity of the
following steps:
\begin{enumerate}
\item The lower bound of the time integral in 
\refeqs{Feom}{rhoeom} is extended to $-\infty$.
The equations are often formulated in Wigner transformed form. In order
to make use of some mathematical identities to rewrite the equations in
a handy form one usually extends the finite integrals in the relative
($x-y$) coordinate to the range $-\infty\dots\infty$. If one does this,
the equations do not refer to an initial value problem any more. 
This introduces an approximation, that 
is justified from that time $t$ on, when $F(t,t_0)$ and $\partial_t\rho(t,t_0)$
become small compared to the equal time correlators.
This usually occurs around damping time ($t_{\rm damp}$),
given by the imaginary part of the self energy \cite{trans}.
\item Transport equations typically use gradient expansion to make the
equations of motion local in time. It is controlled by the smoothness of
the evolution, and becomes exact near equilibrium. As expected, in weakly
coupled theories subsequent orders of this expansion converge well. The
intermediate time description of evolution can be substantially improved
by going beyond the usually considered lowest order \cite{trans}.
\item The transport equation for $\rho$ can be replaced by an assumed
relationship between the $F$ and $\rho$ propagators. This nonequilibrium
KMS condition parametrized by the particle distribution can be used as
an ansatz in the $F$ equation. It has been shown using 2PI equations that this
relationship is dynamically established in damping time scale, well
before the equilibration of the parametrizing distribution \cite{sewm04}.
\item In the final step one replaces the spectral function ($\rho$)
by a delta function, and introduces the quasi-particles. A detailed
comparison of Boltzmann equation restricted to binary collisions with
the 2PI dynamics based on \refeq{2piphi4} can be found in Ref.~\cite{Markus}.
\end{enumerate}
In summary, these steps are well justified in sufficiently weakly coupled
theories, only after the damping time scale. This sets a clear limitation to
transport and kinetic theories when they are applied to initial value problems
\cite{trans}.

\section{Testing 2PI}

\subsection{Convergence in equilibrium}

Since nonequilibrium description of fields requires non-perturbative treatment
we were forced to use a resummation formalism.
The same techniques can be applied in thermal equilibrium, too,
where efficient formulations in Euclidean space-time become available.
In contrast to the far-from-equilibrium case, there are various powerful
approximation schemes known in thermal field theory.
A prominent approach in equilibrium high-temperature
field theory is the so-called ``hard-thermal-loop''
resummation~\cite{Braaten:1989mz}. However, explicit
calculations of thermodynamic quantities such as pressure or entropy typically
reveal a poor convergence except for extremely small couplings. An important
example for this behavior concerns high-temperature gauge theories.  Recent
strong efforts to improve the convergence aim at connecting to available
lattice QCD results, for which high temperatures are difficult to achieve. In
order to find improved approximation schemes it is important to note that the
problem is not specific to gauge field theories. Indeed it has been documented
in the literature in great detail that problems of convergence of perturbative
approaches at high temperature can already be studied in simple scalar
theories. For recent reviews in this context see Ref.~\cite{Blaizot:2003tw}.

A promising candidate for an improved convergence behavior is the loop or
coupling expansion of the 2PI effective action. So far, thermodynamic
quantities such as pressure or entropy have been mainly calculated to two-loop
order. However, aspects of convergence can be sensefully discussed only beyond
two-loop order since the one-loop high-temperature result corresponds to the
free gas approximation.  Efforts to calculate pressure to nontrivial order 
include
so-called approximately self-consistent approximations~\cite{Blaizot:1999ip},
as well as estimates based on further perturbative expansions in the coupling
and a variational mass parameter~\cite{Braaten:2001vr}. These studies indicate
already improved convergence properties. However, perturbatively motivated
estimates as in Ref.~\cite{Braaten:2001vr} suffer from the presence of
nonrenormalizable, ultraviolet divergent contributions and the apparent
breakdown of the approach beyond some value for the coupling.  If one does not
want to rely on these further assumptions, going beyond two-loop order requires
the use of efficient numerical techniques. Such rigorous studies are important
to get a decisive answer about the properties of 2PI expansions. As it turns
out these problems appear as an artefact of the additional
approximations employed and cannot be attributed to the 2PI loop expansion.

In Ref.~\cite{renormthermo} we calculate the renormalized thermodynamics
of a scalar $\Phi^4$ in the 2PI loop expansion to next-to-leading order.
We solve the Schwinger--Dyson equation for the propagator, as well as
the Bethe--Salpeter equation for the four-point function. By adding counterterms
to the 2PI effective action we require the finiteness of the propagator
following Refs.~\cite{vanHeesRenorm,UrkoRenorm}. In addition to this, 
we also renormalize the value of the effective action itself by an additional
counterterm $\sim\Phi^4$. As it has been shown in Ref.~\cite{Julien}, all
$n$-point functions are finite, and all counterterms are temperature
independent.

\begin{figure}
\begin{center}
\includegraphics[width=16cm]{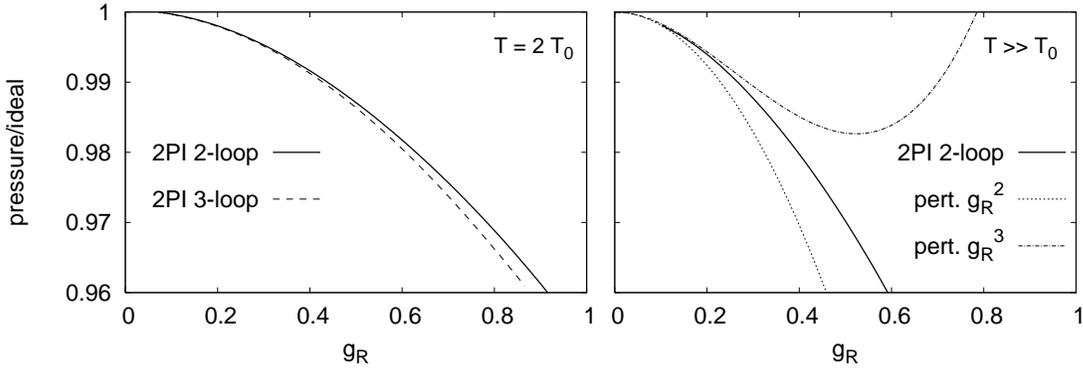}
\end{center}
\caption{\label{fig:cmp2pi}
Convergence of the 2PI resummation (left) and the perturbative series (right)
in equilibrium. The scalar $\Phi^4$ theory has been renormalized at a
temperature equal to the thermal mass $T_0=m_R(T_0)$, and also solved at $T$
for which the pressure is shown. The subsequent orders of 2PI show a good
convergence behavior even at higher couplings \cite{renormthermo}.}
\end{figure}

The bad convergence behavior of the perturbation theory is \textit{not} seen
in the 2PI results (\reffig{fig:cmp2pi}).
To show this, we calculate the pressure to leading
and next-to-leading order in the loop expansion, both in the 2PI and 
in the standard perturbative expansion. We could explore a broad range
in the coupling, going very close to the triviality limit.

\subsection{Testing out of equilibrium}

Although the results from the 2PI formalism seem plausible, it has to be
checked against other reliable nonequilibrium methods.
If the exact evolution of
the quantum theory were available, this test would be straightforward.

In fact, the classical field theory approximation, provides the only
way to solve the full non-perturbative dynamics of a field theory for
arbitrary late times. In the classical limit the operator equations
reduce to wave equations, which are convenient for numerical treatment. 
The classical approach became very popular both for heavy ion physics
\cite{lappi} and in cosmology \cite{khlebnikov}, where the high occupation
numbers keep the applications within the classical limit. The appearance of
Rayleigh--Jeans divergences and the lack of genuine quantum effects, however,
limit their use.

Still, the classical approximation proved to be extremely useful for
benchmarking the 2PI resummation in a nonequilibrium situation. 
The 2PI formalism is a generic field theoretical tool, that can be formulated
both in quantum and in classical context. Aarts and Berges
compared the classical and quantum dynamics of an O($N$) symmetric scalar 
field based on the large $N$ expansion of the 2PI effective action,
and solved the exact classical dynamics as well \cite{AartsBergesClassical}.

\begin{figure}
\begin{center}
\includegraphics[width=8cm]{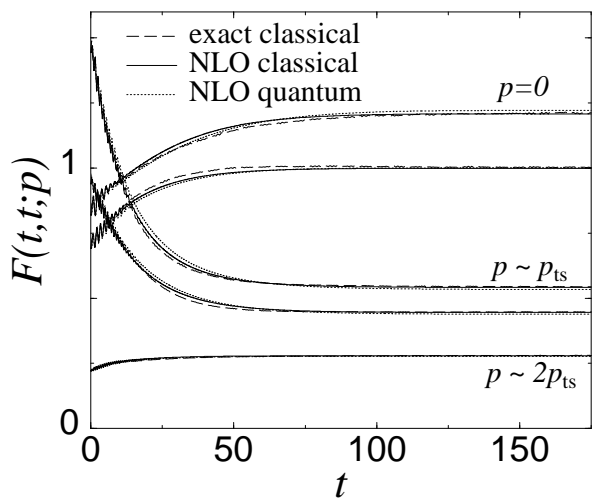}
\end{center}
\caption{\label{fig:ClassicalComparison}
Nonequilibrium evolution of the equal time propagator in a scalar O($N$)
model for various momenta. Modes around $p_{\rm ts}$ were initially
excited and their dynamics has been followed by solving
the exact classical equations and the next-to-leading order 2PI
equations both in the quantum theory and in the classical approximation.
The agreement of the curves for $N=10$ shows that the 2PI techniques provide
a very good approximation scheme even for late times.
\cite{AartsBergesClassical}.
}
\end{figure}

In their comparison (see Fig.~\ref{fig:ClassicalComparison}) for sufficiently
high occupation numbers and large number of fields all the three approximations
gave the same result. Even at lower occupancies and smaller $N$ they found
a good agreement between the classical 2PI and the exact classical results.
This suggests that the selective resummation of the 2PI formalism includes the
essential diagrams for irreversible dynamics. 

Future tests of 2PI dynamics might become available by a recent observation by
Berges and Stamatescu \cite{BergesStamatescu}. They carried out a lattice
simulation in Minkowski space--time by using a reformulation of stochastic
quantization for the path integral \cite{stochquant}. Although not much is
known about the general convergence properties, this simulation technique is a
promising non-perturbative approach to out-of-equilibrium fields. A positive
result of comparing 2PI with the lattice simulation would mark a breakthrough
in nonequilibrium field theory \cite{sim2pi}.

\section{Time scales of nonequilibrium evolution}

\begin{figure}[t]
\begin{center}
\includegraphics[width=8.5cm]{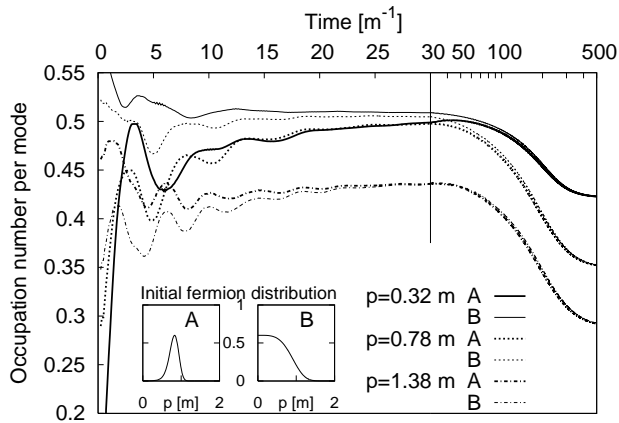}
\end{center}
\vspace*{-0.7cm}
\caption{Fermion occupation number 
for three different momentum modes as a function of time
in the chiral quark meson model of Ref.~\cite{Fermions}.
} 
\label{fig:join_fn}
\end{figure}

The 2PI equations of motion have been in use for several years, already.
Most applications focused on thermalization \cite{CoxBerges,
AartsBergesSpectral,BergesON,prethermalization,Fermions,
Arrizabalaga1,Arrizabalaga2,Isotropization,trans} i.e. the gradual loss of all
initial information but the energy density. All these applications confirmed
that different quantities effectively thermalize 
on different time scales. Hence, a partial thermalization may be sufficient
to support the assumptions of thermal approaches in cosmology or in
heavy ion physics. 
This has been pointed out in Ref.~\cite{prethermalization}, where it was shown 
for a chiral quark-meson model that the prethermalization of important 
observables occurs on time scales dramatically shorter than the thermal 
equilibration time. As an example, Fig.~\ref{fig:join_fn} shows 
the nonequilibrium time 
evolution of fermion occupation number for three different momentum modes
in this model. The evolution is given for two different initial
particle number distributions A and B shown in the insets, 
with {\em same} energy density. The vertical line marks
the characteristic time scale $\sim t_{\rm damp}$, after which 
the details about the initial distributions A or B are effectively lost.
The following long-time behavior to thermal equilibrium is shown on a 
logarithmic scale in units of the scalar 
thermal mass $m$.

\begin{figure}[t]
\begin{center}
\includegraphics[width=8.5cm]{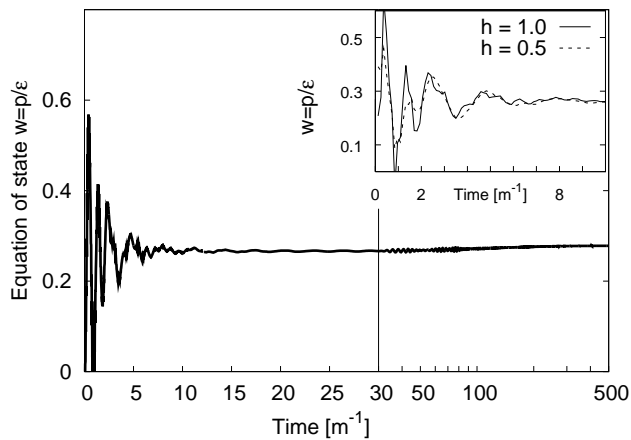}
\end{center}
\vspace*{-0.7cm}
\caption{The ratio of average pressure over energy density $w$ 
as a function of time. The inset shows the early stages 
for two different couplings $h$ and demonstrates
that the prethermalization
time is rather independent of the interaction details.}
\label{fig:wevol}
\end{figure}

In contrast to the very long time $t_{\rm eq}$ for 
complete thermal equilibration,
prethermalization of the (average) 
equation of state sets in extremely rapidly on a time scale
\begin{equation}
t_{\rm pt} \ll t_{\rm damp} \ll t_{\rm eq} \, .
\end{equation}
In Fig.~\ref{fig:wevol} 
we show the ratio of average pressure (trace over space-like components
of the energy-momentum tensor) over energy density,
$w = p/\epsilon$, as a function of time. One observes that
an almost time-independent equation of state builds up 
very early, even though the system is still far from equilibrium!
Here the prethermalization time $t_{\rm pt}$ is 
of the order of the characteristic
inverse mass scale $m^{-1}$. This is a typical consequence of
the loss of phase information by summing over oscillating functions 
with a sufficiently dense frequency spectrum.
If the ``temperature'' ($T$), i.e.~average kinetic energy per mode,
sets the relevant scale one finds  
$T\, t_{\rm pt} \simeq 2 - 2.5$~\cite{prethermalization}.
For $T \gtrsim 400 - 500\,$MeV one
obtains a very short prethermalization
time $t_{\rm pt}$ of somewhat less than $1\,$fm.

This is consistent with very early hydrodynamic behavior, however,
it is not sufficient as noted in Refs.~\cite{prethermalization,Arnold:2004ti}.
Beyond the average equation of state, a crucial ingredient
for the applicability of hydrodynamics for collision 
experiments~\cite{Heinz:2004pj} is the approximate isotropy 
of the local pressure. More precisely, the diagonal (space-like) components
of the local energy-momentum tensor have to be approximately equal.
Of particular importance is the possible isotropization far from 
equilibrium. The relevant time scale for the early validity 
of hydrodynamics could then be set by the isotropization 
time. The analysis of scalar models lead to an isotropization
time given by the comparably long characteristic damping time 
$\sim t_{\rm damp}$ \cite{Isotropization}. In gauge theories, however,
there is a weak-coupling mechanism for faster isotropization
identified as plasma instabilities
\cite{Arnold:2004ti,plasmainstabilities1,plasmainstabilities2}.
Whether this can explain
the observations or whether they suggest that we have 
to deal with some new form of a ``strongly coupled Quark Gluon Plasma''
is an important open question. 

\subsection*{Acknowledgments}
The speaker acknowledges the collaboration with
J\"urgen Berges, Urko Reinosa, Julien Serreau and Christof Wetterich
on related subjects and the fruitful discussions with
%Gert Aarts and
Antal Jakov\'{a}c.

\end{document}